\documentclass[letterpaper,twocolumn,10pt]{article}
\usepackage{usenix2022_SOUPS}

\usepackage{tikz}
\usepackage{amsmath}

\usepackage{filecontents}
\usepackage{color} 
\usepackage{booktabs}
\usepackage{makecell}
\usepackage{lipsum}
\usepackage{amssymb}

\begin{document}

\date{}

\title{\Large \bf New Differential Privacy Communication Pipeline and Design Framework}

\def\plainauthor{Author name(s) for PDF metadata. Don't forget to anonymize for submission!}

\author{
{\rm Jingyu Jia}\\
Nankai University
\and
{\rm Zikai Alex Wen\footnotemark[1]} \\
Hong Kong University of \\ Science and Technology \\ (Guangzhou)
\and
{\rm Zheli Liu}\\
Nankai University
\and
{\rm Changyu Dong\footnotemark[1]}\\
Guangzhou University
} 

\maketitle

\begin{abstract}
Organizations started to adopt differential privacy (DP) techniques hoping to persuade more users to share personal data with them. However, many users do not understand DP techniques, thus may not willing to share. Previous research suggested that the design of DP mechanism communication could influence users' willingness to share data. Based on the prior work, we propose a new communication pipeline that starts with asking users about their privacy concerns and then provides customized DP mechanism and communication. We also propose a design framework that systemically explores effective communication designs ranging from a text-based high-level description to a step-by-step interactive storyboard. Based on the framework, we created 17 designs and recruited 5 people to evaluate. Our user study showed that text-based descriptions have the highest clarity in all scenarios, while the step-by-step interactive storyboards have potential in persuading users to trust \textit{central DP}. Our future work will optimize the design and conduct a large-scale efficacy study.
\end{abstract}

\renewcommand{\thefootnote}{\fnsymbol{footnote}}
\footnotetext[1]{Zikai Alex Wen and Changyu Dong are the corresponding authors.}
\renewcommand{\thefootnote}{\arabic{footnote}}
\section{Introduction}
Organizations have been striving hard to persuade users to share personal data with them. To achieve this goal, the organizations must use privacy-preserving technologies to prevent users' sensitive data from being leaked. In addition, they need to convince their users to trust that their technologies can indeed protect users' privacy, so that the users will be willing to share data. That being said, many technologies are available for data protection. To identify what privacy-preserving technology is suitable for a particular occasion, there are needs for proofing and comparing the data protection efficacy of different technologies. To meet the needs, security and privacy researchers~\cite{Dwork2006, Dwork2014, Erlingsson2014, Bittau2017} have been working on a technique called differential privacy (DP).

\begin{table}[!htbp]
\renewcommand\arraystretch{1.5}
\centering
\footnotesize
\caption{Seven Types of Privacy Concerns}
\label{table:concerns}
\begin{tabular}{p{0.01\textwidth} p{0.07\textwidth} p{0.34\textwidth}}
\toprule
No. & Abbreviation & Description \\ \hline
1 & Hack & My data will be hacked by hackers.\\
2 & Law & My data will be forcibly acquired by the government.\\
3 & Organization & My data will be stolen by unrelated employees in the organization.\\
4 & Disclosure & My data will be disclosed to others by the organization. \\
5 & Analyst & My data will be accessed by the data analysts in the organization. \\
6 & Graphs & The graphs and tables generated by the organization will reveal my data. \\
7 & Share & The organization will reveal my data when sharing the processed dataset with others.\\
\bottomrule
\end{tabular}
\end{table}

\begin{figure}[!h]
    \centering
    \vspace{-3mm}
    \includegraphics[width=0.5\linewidth]{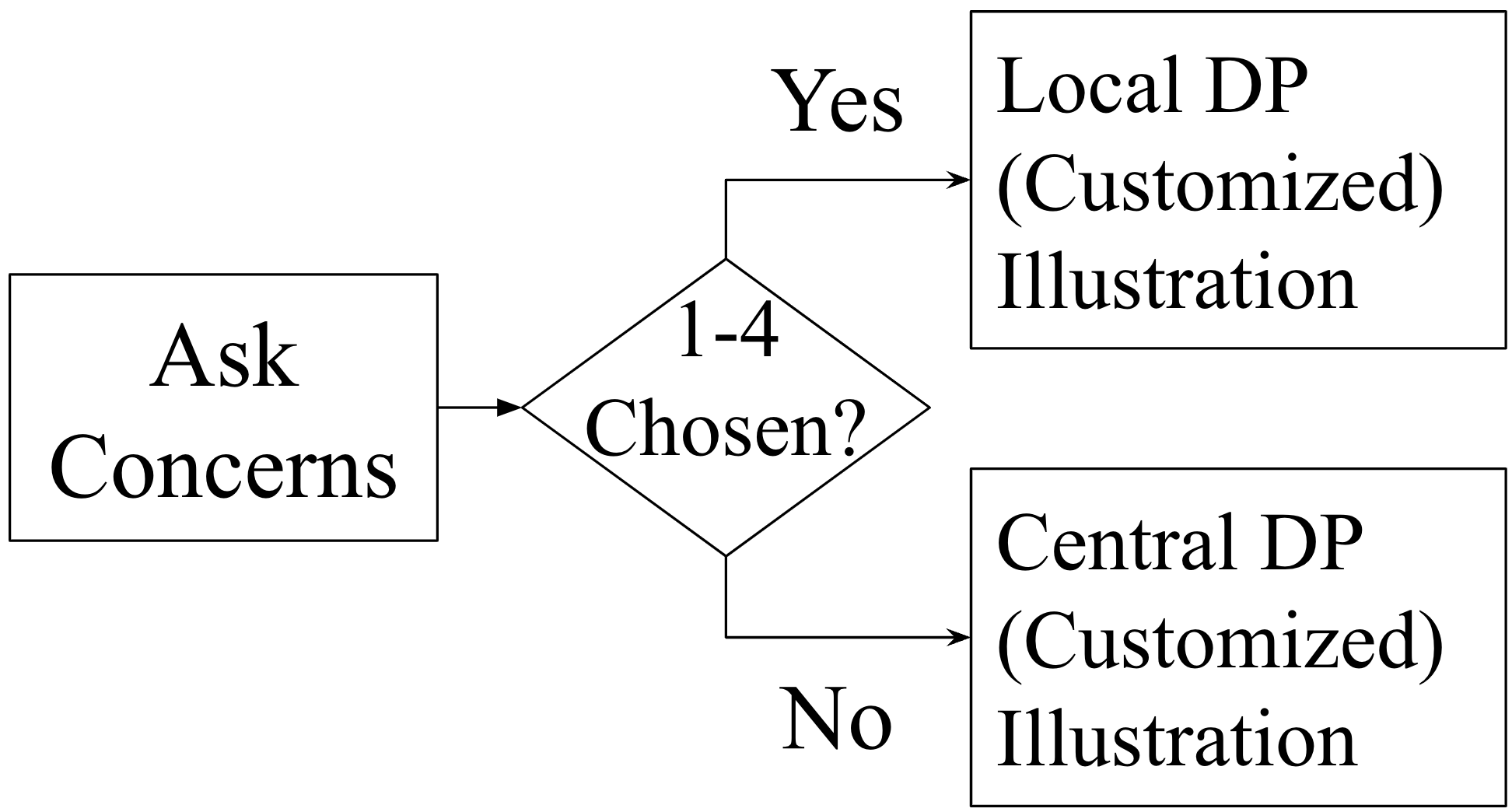}
    \vspace{-3mm}
    \caption{The communication pipeline asks about the user's privacy concerns to match customized DP and illustration.}
    \label{fig:pipeline}
\vspace{-3mm}
\end{figure}

DP evaluates the efficacy of a privacy-preserving technology by quantifying the difference between two data analysis results that are processed by the technology. These two analysis results are calculated from two datasets: one dataset that collects the individual user's data and another dataset that does not collect the data. If the aforementioned results are not significantly different, then the privacy-preserving technology is proved to satisfy the DP definition. In this case, it is unlikely to identify an individual user's record through queries protected by a DP mechanism.

However, it is difficult for ordinary users to understand the proof of DP. Therefore, the organizations also need to persuade users that using DP can prevent their personal data from being leaked. Currently, most organizations rely on text-based descriptions to communicate the DP mechanism with users (e.g., Apple's white paper~\cite{Apple2017}). In addition, it is still debatable whether the existing DP communication designs, including text-based~\cite{Cummings2021} and animation-style~\cite{Xiong2022} designs, may persuade more users to share their data.

Therefore, our work takes a step further to seek effective ways to communicate DP mechanisms with ordinary users. First, we combined key takeaways from prior work~\cite{Cummings2021, Xiong2022} to design a new communication pipeline (as shown in Figure~\ref{fig:pipeline}) that attempts to address different types of user concerns. Second, we proposed a communication design framework that consists of four categories: (1) text-based description, (2) data input/output illustration, (3) data output probability distribution illustration, and (4) step-by-step interactive storyboards.

To study what communication design would be clear and persuasive, guided by the framework, we created nine designs to persuade users into sharing \textit{numerical} salary data and eight designs to persuade users into sharing \textit{geographical} location data (as shown in Figure~\ref{fig:designs}). After that, we conducted a user study with four ordinary users and one DP researcher to evaluate the 17 designs and provide feedback about narrowing the designs down to the effective ones. We list and discuss our key findings in the user study section.

\vspace{-3mm}
    
\section{Related Work}

Recently, an increasing number of research projects have studied how to help users understand DP mechanism and to convince users to share their data~\cite{Cummings2021,Karegar2021,Xiong2020,Xiong2022,Nanayakkara2022,Franzen2022}. Xiong et al.~\cite{Xiong2020, Xiong2022} investigated how textual and illustration descriptions of DP affect users' understanding of DP and willingness to share data. They found that providing users with DP descriptions facilitated data sharing and that illustrations effectively describe DP models to users. Nanayakkara et al.~\cite{Nanayakkara2022} designed an interactive visualization tool for data managers who are experienced in sensitive data analysis but unfamiliar with DP to help them set effective DP parameters. 

Franzen et al.~\cite{Franzen2022} explained DP to users through a risk communication format, and they found that emphasizing DP risks rather than DP functionality, while not affecting users' objective understanding of DP, can reduce subjective confidence in their understanding. Cummings et al.~\cite{Cummings2021} found that in-the-wild DP descriptions neither addressed users' privacy concerns nor promoted data sharing. They proposed that descriptions should be tailored to address specific user privacy concerns to promote user data sharing. Based on the previous work, our goal is to design a new DP communication pipeline and interactive DP mechanism illustrations that can be customized to address different user privacy concerns, and thus facilitating data sharing.
\section{New Pipeline and Design Framework}
Our work aims to find effective ways to explain to users how their data is protected by a DP technique. We hope that our work would change people's mind if they do not want to share data because they do not understand DP. We deduced from Cummings et al's research~\cite{Cummings2021} that users would prefer a customized privacy notification to address their specific concerns. Therefore, we propose to change the traditional ``read-then-consent'' approach to inform users the privacy policy. We designed a new communication pipeline that starts with asking users their privacy concerns then provides the suitable DP technique and a customized privacy notification. Our work focuses on designing a critical part of the privacy notification: a clear and persuasive DP technique explanation. To systemically study what designs meet our goal, we propose a framework that covers four design categories ranging from non-interactive high-level description to interactive storyboard illustration.

In this section, we first describe the design of our DP communication pipeline and explain the rationale behind the design. Then, we describe the framework for designing DP communication. Following this, we describe how we followed the design framework to create nine designs for sharing salary data and eight designs for sharing location data.

\textbf{Communication Pipeline.} As shown in Figure~\ref{fig:pipeline}, the communication pipeline starts from presenting seven major data privacy concerns to the user then asks the user to choose the concerns that they have. The definition of seven privacy concerns (as shown in Table~\ref{table:concerns}) is adapted from the definition proposed by Cumming et al.~\cite{Cummings2021}. We added \textit{Disclosure} to the concerns to provide finer granularity, which allows the system to identify how much trust the user has in the organization. 

If it concerns the user that the organization may disclose raw user data without their additional consent then they should choose \textit{Disclosure}. But if the user trusts the organization to share privacy-preserving data with others and only have concern about the reliability of the privacy-preserving technique, then they can choose \textit{Share}. We also refined the description of \textit{Organization} so that if the user worries about unauthorized data access inside the organization then they should choose it. If the user worries about authorized data access inside the organization, then they can choose \textit{Analyst}.

\begin{figure*}[!h]
    \centering
    \includegraphics[width=\linewidth]{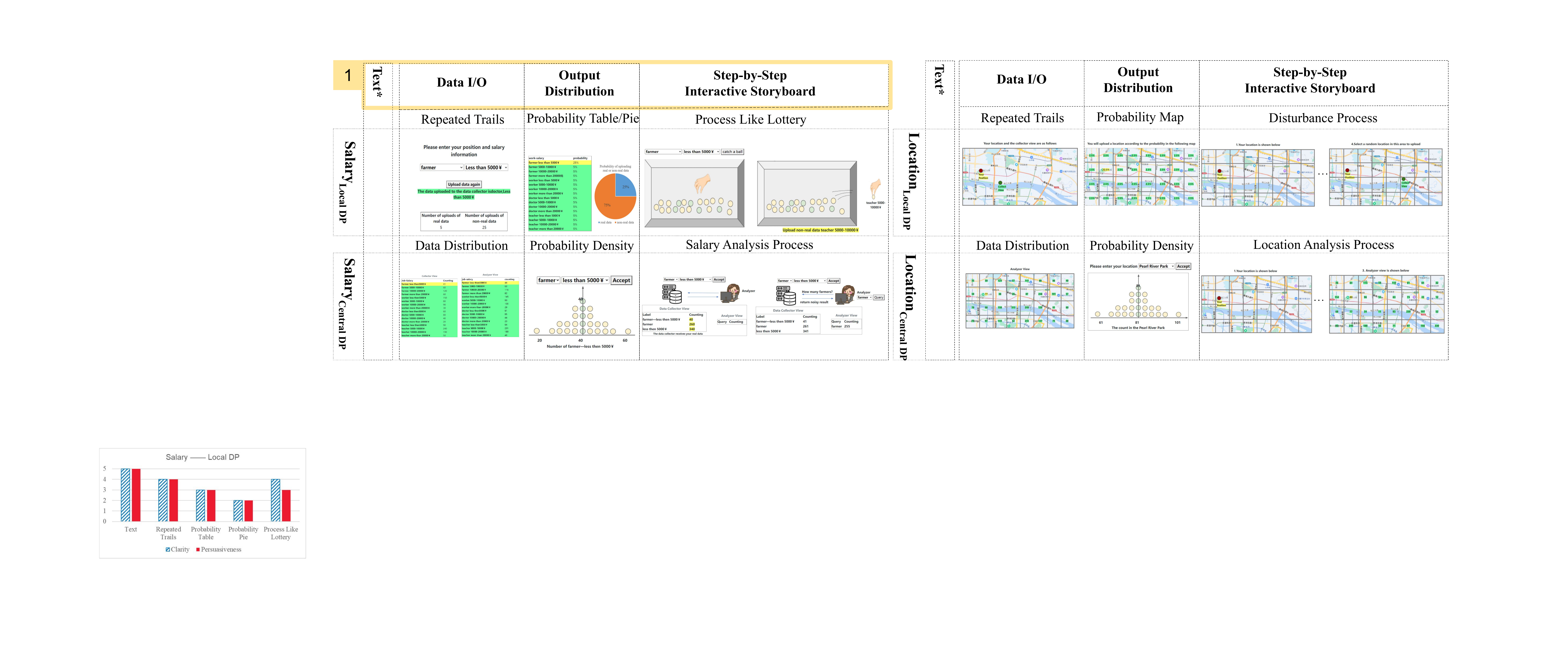}
    \vspace{-2em}
    \caption{This diagram shows our design framework of four design categories. \textit{Highlight 1} shows the four design categories. It also shows nine designs for salary data request scenario and eight designs for location data request scenario. \textit{Text*} is the category of all text-based descriptions for four design scenarios: \textit{local DP} and \textit{central DP} for salary data protection and \textit{local DP} and \textit{central DP} for location data protection. The designs are available online\textsuperscript{\ref{git}}.}
    \label{fig:designs}
\end{figure*}

\begin{figure*}[!h]
    \centering
    \includegraphics[width=\linewidth]{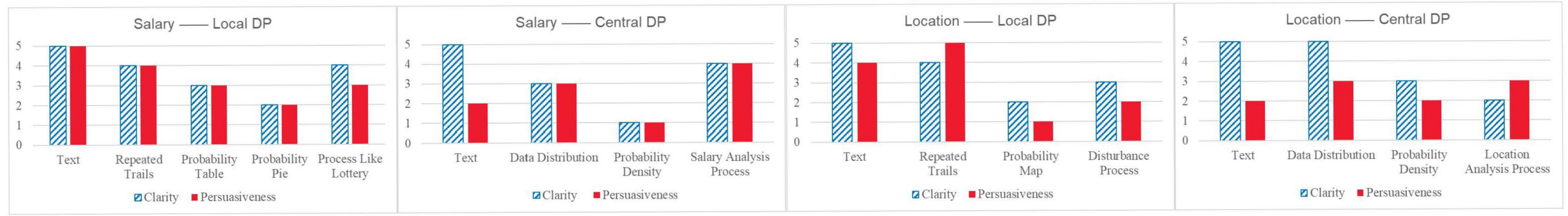}
    \vspace{-2em}
    \caption{The histograms show the number of participants agreed or strongly agreed that a design is clear or persuasive.}
    \label{fig:histogram}
\end{figure*}

After the system receives the user's response, it decides which level of DP, \textit{local DP} or \textit{central DP}, can address the user's concerns. Then the system explains to the user how the DP mechanism would resolve their concerns before it asks for consent. According to our analysis, if the user chooses one of the concerns from No.1 to No.4, then the safest way is to not pass real data to the organization. In this case, they need the strict DP mechanism: \textit{local DP} that adds noise to the data on the user's local computing platform before sending the data to the organization. In comparison, if the user only chooses the concerns from No.5 to No.7, then the system can use a less strict DP mechanism: \textit{central DP} because the user indicates that they trust the organization to protect their privacy. By using \textit{central DP}, the organization can store users' raw data and only adds noise to the original data when sharing it with authorized parties to preserve users' privacy.

\textbf{Design Framework.} Now that we have the ability to understand a user's specific privacy concerns through the communication pipeline, our next job is to design a customizable DP mechanism explanation that can allay different types of user concerns. To systemically study possible effective designs, we propose a framework that consists of four design categories (as shown in Figure~\ref{fig:designs}, Highlight 1): (1) text-based description, (2) data input/output illustration, (3) data output probability distribution illustration, and (4) step-by-step interactive storyboard illustration.

We created this framework to cover a spectrum of design categories: Category (1) provides a high-level description of a DP mechanism; Categories (2) and (3) provides pictorial illustrations about data output values or probability distributions; and Category (4) is a new type of design that we would like to explore: designing storyboards to explain a DP mechanism and using step-by-step interactions to scaffold the understanding. The design complexities increase from categories (1) to (4), but we suspect that a design is more user-friendly if it is contextualized and provides step-by-step interactive scaffolding.

To validate our hypothesis, based on the design framework, we created 17 designs for two data sharing request scenarios: nine designs for requesting salary data and eight designs for location data. Previous work~\cite{Cummings2021, Xiong2022} designed communication methods for these two scenarios respectively. Cumming et al.'s design~\cite{Cummings2021} falls into category (1) and Xiong et al.'s design~\cite{Xiong2022} belongs to category (2-3) and the non-interactive version of (4). The following paragraphs describe the details of all 17 designs in a 4x4 design space: four design categories times four types of DP protected data sharing scenarios (i.e., protect salary data using \textit{local DP} or \textit{central DP} and protect location data using \textit{local DP} or \textit{central DP}).

\textit{(1) Text-Based Description:} We followed the design of text-based description by Cummings et al.~\cite{Cummings2021} to cover three pieces of DP information: (1) the specific data to be collected, (2) the stage to disturb the collected data, and (3) the explanation about why the party that concerns the user cannot infer personal data. We designed a text-based description template for each DP protected data sharing scenarios. The template includes seven sentences that aims to allay the seven privacy concerns respectively. These sentences can be formed into one paragraph if needed. Due to the page limited, we posted the templates document online~\footnote{\label{git}\url{https://github.com/zikaiwen/SOUPS22}}. We treat this design category as a design baseline because mainstream organizations (e.g., Apple~\cite{Apple2017}) are using this approach to explain the DP mechanism to their target users.

\textit{(2) Data Input/Output Illustration:} As shown in Figure~\ref{fig:designs}, we designed two \textit{Repeated Trails} to illustrate the output processed by \textit{local DP} and two \textit{Data Distribution} for \textit{central DP}. After the user enters an example data input, \textit{Repeated Trails} shows the data output that the organization obtains. By repeatedly providing an example input, users may understand that the data output keeps changing under a certain rule so it may protect their privacy. \textit{Data Distribution} illustrations showcase what data output the organization can produce for data queries. Users may understand that their individual records are accumulated. And even the accumulated dataset does not reflect the true user inputs. These designs illustrate data outputs, but they do not illustrate how the outputs are produced.

\textit{(3) Data Output Probability Distribution Illustration:} We adapted Nanayakkara et al.'s design of discrete visualization of the statistical density graph~\cite{Nanayakkara2022} to create \textit{Probability Density} illustrations. We followed their design because the design could help ordinary data managers understand the distribution of DP protected data. Managers can calculate the proportion of balls to the left of the output value to learn the probability of getting such an output value. We also followed the traditional practice to design \textit{Probability Table and Pie Chart} and \textit{Probability Map} to illustrate the mechanism of \textit{local DP}. We suspect that these designs may require users to have knowledge about probability and statistics to comprehend. 

\textit{(4) Step-by-Step Interactive Storyboard Illustration:} The illustrations in this category may look various. However, we created them following certain guidelines. To illustrate how \textit{local DP} functions, we designed storyboards that animate how the data input is disturbed on the local computing platform before it is obtained by the organization. To illustrate how \textit{central DP} functions, we designed storyboards that first animate how the input data is obtained by the organization without any disturbance, then animate what the data queries can be, and finally animate how the organization adds noise to the data before it answers the data query. We suppose the storyboard can provide context and the step-by-step interactions provide mind breaks, which may make users feel more intuitive and easy to follow.
\section{Evaluation Study}
To verify what DP communication designs are clear to ordinary users, and more importantly, can persuade users to share their data under DP protection, we conducted a user study with five people. Our user study received the ethical approval of Guangzhou University.

\textbf{Participants.} We recruited five participants (4 ordinary users and 1 DP expert) through email and social media. The ordinary users' ages ranged from 19 to 57 years old. Two of them got high school degrees and the other two got bachelor degrees. The DP expert is 31 years old and got a Ph.D. degree. They had between 9 and 30+ years of experience in using Internet. All of them experienced privacy breach and encounter bad consequences including harassment and/or receiving spam or phishing messages.

\textbf{Procedure.} Participant started by learning about DP knowledge from us to ensure that they have sufficient background knowledge to evaluate the designs and express their thoughts during the study. This learning process is not included in the final communication design. After the participant finished learning, we asked them to experience the privacy communication pipeline and designs on our crafted websites: One is an income justice study survey and the other is a restaurant recommendation system. During the study, the participant was asked to answer two five-point likert-scale questions for each design: (1) This design clearly describes the differential privacy mechanism; and (2) This design resolves my concern(s) so I feel comfortable to share my data. We followed up on the participant's evaluation and prompted them to refine the design if they have any thoughts. The entire study took about one and a half hours.

\textbf{Method.} We conducted the study face-to-face or remotely over Zoom. Participants received \yen 200 Chinese Yuan in cash. All user studies were audio and screen recorded then transcribed. We used an open-coding technique~\cite{Saldana2021} to find common themes shared across the studies. In the key findings, we include quotes from participants identified by ID numbers following the letter P for “participant” (e.g. P1).

\textbf{Key Findings and Discussion.} All participants agreed or strongly agreed that text-based descriptions are clear under all conditions. That being said, \textbf{text-based description may not be the best approach to explain a \textit{central DP} mechanism}. While all participants agreed or strongly agreed that texts were clear, only two participants were persuaded to share data. Let us take the location data request scenario for instance. Among them, P5 said they were willing to share even if the description was not clear because they had been sharing their location data all the time. In contrast, P2 was extremely cautious about sharing their location information because they had a terrible harassment experience after their location information was leaked. So, they rated the persuasiveness of all designs for \textit{central DP} $\mathrm{\leqslant 3}$. The rest three participants depend on the quality of communication material to decide whether they share location data. Text-based description was the least persuasive to them, while step-by-step storyboard illustrations showed potential in increasing the persuasiveness as long as it was clear to the user. P4 suggested that the text-based description was clear about how \textit{central DP} mechanism can resolve his No. 7 concern, but they were skeptical until they could contextualize how their input became a privacy-preserving output as they interacted with the storyboard. So, our next design is to \textbf{combine text-based description with step-by-step interactive storyboard illustration}.

\section{Conclusion and Future Work}
In summary, we propose an new communication pipeline to first ask about the user's privacy concerns before providing a privacy notification. When it comes to designing the privacy notification, we propose to present a customized text-based description combined with a step-by-step interactive storyboard to illustrate how the DP mechanism may resolve concerns. We will implement it and conduct an ``in-the-wild'' experiment to test the efficacy of our proposed design.


\bibliographystyle{plain}
\bibliography{ref}

\end{document}